\documentclass[aps,preprint]{revtex4}
\usepackage[dvips]{graphics,graphicx}

\begin{document}
\title{Cyclotron-Bloch dynamics of a quantum particle in a 2D lattice}
\author{Andrey R. Kolovsky$^{1,2}$}
\affiliation{$^1$Kirensky Institute of Physics, 660036 Krasnoyarsk, Russia}
\affiliation{$^2$Siberian Federal University, 660041 Krasnoyarsk, Russia}

\author{Giorgio Mantica$^{3,4}$}
\affiliation{$^3$Center for Nonlinear and Complex Systems, 
University of Insubria, 22100 Como, Italy} 
\affiliation{$^4$CNISM unit\`a di Como, and  I.N.F.N. sezione di Milano, Italy}
\date{\today}

\begin{abstract}
This paper studies the quantum dynamics of a charged particle in a 2D square lattice, under the influence of electric and magnetic fields, the former being aligned with one of the lattice axes and the latter perpendicular to the lattice plane. While in free space these dynamics consist of uniform motions in the direction orthogonal to the electric field vector, we find that, in a lattice, this directed drift takes place only for specific initial conditions and for electric field magnitudes smaller than a critical value. Otherwise, the quantum wave--packet spreads ballistically in both directions orthogonal to the electric field. We quantify this ballistic spreading and identify the subspace of initial conditions insuring directed transport with the drift velocity. We also describe the effect of disorder in the system.
\end{abstract}
\maketitle

\section{Introduction}
Quantum transport in periodic potentials is a topic of permanent interest since the early days of quantum mechanics. Recently this topic has attracted a renewed interest thanks to the experiments with cold atoms in optical lattices, where the quantum dynamics can be observed both in real and in momentum space. In particular, during the last decade much attention has been devoted to the phenomenon of Bloch oscillations of cold atoms subject to a static (for example, gravitational) field (see {\em e.g.} Ref.~\cite{Daha96,Mors01,80,Hall10} to cite just a few of the relevant papers). These new studies have also shed additional light on the old problem of electron transport in a solid crystal induced by an external electric field \cite{Ott04,69,79}.

Present research in cold atoms physics is also focused on the problem of generating synthetic magnetic fields, which could impart a Lorenz--like force to otherwise neutral atoms in motion \cite{Jaks03,Oste05,Sore05,preprint}. This opens an interesting perspective for studying the quantum Hall effect with cold atoms in 2D optical lattices \cite{Gold07}. A preliminary but necessary step in this direction is the analysis of the Bloch dynamics of cold atoms in the presence of an artificial magnetic field \cite{remark0}. In this paper we study these dynamics in the tight-binding approximation, where the single-particle Hamiltonian of an atom in a 2D lattice has a relatively simple form. In spite of this simple form, we find a surprisingly rich variety of behaviors, which range from the ballistic spreading of the wave-packet to directed transport.

The structure of the paper is as follows. In Sec.~\ref{sec2} we introduce the model and perform a preliminary analysis, following the standard route which leads to the Harper equation \cite{Harp55} and, hence, is referred to as  the Harper approach. The Harper approach reduces the original 2D problem to a 1D problem, which on the one hand is simpler to describe, but on the other hand it limits the class of initial conditions that can be considered. Therefore, in Sec.~\ref{sec3} we tackle the problem using the more powerful formalism of Landau-Stark states, which are the eigenstates of a charged particle in a lattice subject to both magnetic and electric fields. In some sense this approach is opposite to the Harper approach: here one begins with a particle in an electric field (i.e., with the problem of Bloch oscillations) and then introduces a magnetic field. The Landau-Stark states approach is fully two dimensional and, hence, imposes no limitations on the class of initial conditions that can be considered. In addition, it allows us to construct 2D localized wave packets, which propagate across the lattice without changing their shapes.

Finally, in Sec.~\ref{sec4} we discuss the transport in the presence of on-site disorder. Here we consider generic initial conditions in the form of a wide incoherent wave packet. In this case, for vanishing disorder, the wave--packet spreads ballistically in the direction orthogonal to the electric field. We study the modifications of this ballistic spreading caused by the presence of on-site disorder. In the case of null magnetic field this problem was considered in the recent paper \cite{78}. Our analysis unveils the differences between the 1D Harper approach of Sec.~\ref{sec2} and the full 2D approach of  Sec.~\ref{sec3}, that, to the contrary, predict the same rate of ballistic spreading when disorder is absent or negligible. This provides us with a significant new model for the problem of Anderson localization, where the effective dimensionality is larger than one but smaller than two.

\section{The Hamiltonian model and its approximations}
\label{sec2}
In this section we introduce the physical model under investigation and we discuss two common approximations. We also derive a first dynamical phenomenon: the presence of a drift of the particle/wave--packet with constant velocity. Let us therefore consider a quantum particle of mass $M$ and charge $e$ in a 2D square lattice of side $d$ in the $x$-$y$ plane, created by a periodic potential $V(x,y)$. The particle is also under the action of an in-plane electric field, aligned with the $y$ axis and of a magnetic field normal to the $x$-$y$ plane. This particle is described by the Hamiltonian
\begin{equation}
\label{1}
\widehat{H}=\frac{1}{2M}\left(\hat{\bf{p}}-\frac{e}{c}{\bf A}\right)^2+V(x,y) \;,\quad
V(x+ld,y+md)=V(x,y) \;,
\end{equation}
where ${\bf A}$ is the vector potential. A convenient choice of the vector potential, corresponding to the field configuration described above is
\begin{equation}
\label{1a}
{\bf A}=B(0,x,0)+F(0,ct,0) \;,
\end{equation}
where $B$ and $F$ are the magnetic and electric field magnitudes, respectively, and where vectors are described in line notation. It is easy to check that the Hamiltonian (\ref{1}) commutes with the generalized translation operator $\widehat{T}'_x=\exp(-i2\pi\alpha y/d)\widehat{T}_x$, 
where $\widehat{T}_x$ is the usual translation operator and where
\begin{equation}
\label{2}
\alpha=\frac{eBd^2}{hc} \;.
\end{equation}
The parameter $\alpha$ defines the so-called magnetic period $\lambda = d/\alpha$, which has the dimensions of a length and will play an important role in the following. A rigorous analysis of the motion generated by the Hamiltonian (\ref{1}) is a difficult, unsolved problem. Therefore, one deals with this problem by introducing suitable approximations.

\subsection{The effective mass approximation}
\label{sec2a}
The simplest approximation of the Hamiltonian (\ref{1}) starts by defining the effective mass $M^*$ as the curvature of the ground Bloch band of (\ref{1}) at ${\bf A}=0$:
\begin{equation}
\label{3}
M^*=(M_x M_y)^{1/2} \;,\quad M_{x,y}=\frac{1}{\hbar^2}\frac{d^2 E}{d\kappa_{x,y}^2} \;,
\end{equation}
where $E=E(\kappa_x,\kappa_y)$ is the dispersion relation. The approximate Hamiltonian then reads
\begin{equation}
\label{4}
\widehat{H}_{em}=\frac{\hat{p}_x^2}{2M_x}
+\frac{1}{2M_y}\left(\hat{p}_y-\frac{eB}{c}x-eFt\right)^2 \;.
\end{equation}
For vanishing electric field the eigenfunctions of (\ref{4}) are Landau oscillators with energy spectrum $E_n=\hbar\omega_c(n+1/2)$. Here $\omega_c=eB/cM^*$ is the cyclotron frequency. It appears from eq.~(\ref{4}) that a non-zero value of $F$ shifts the origin of a Landau oscillator linearly in time. Thus, the particle is transported in the direction orthogonal to the electric field with a drift velocity $v^*=cF/B$.

Although the effective mass approximation correctly predicts the drift velocity, it completely ignores the lattice discreteness, which is important in many aspects. To take the lattice discreteness into account one usually introduces a tight-binding approximation to the original Hamiltonian.

\subsection{The tight-binding approximation}
\label{sec2b}
The tight-binding approximation amounts to the following ansatz for the wave function of the system,
\begin{equation}
\label{5}
\Psi(x,y)=\sum_{l,m} \psi_{l,m} w_{l,m}(x,y) \;,\quad
w_{l,m}(x,y)=w_{0,0}(x-ld,y-md) \;,
\end{equation}
that is written in terms of two-dimensional Wannier states $w_{l,m}(x,y)\equiv |l,m\rangle$. In this approximation one also neglects coupling other than between nearest neighbors pairs of states, so to approximate the Hamiltonian (\ref{1}-\ref{2}) in the form
\begin{equation}
\label{6}
\widehat{H}_{tb}= -\frac{J_x}{2} \sum_{l,m} \left(|l+1,m\rangle \langle l,m | + h.c.\right)
-\frac{J_y}{2} \sum_{l,m} \left(|l,m+1\rangle \langle m | e^{i(2\pi\alpha l-\omega_Bt)} + h.c.\right)  ,
\end{equation}
where $\omega_B=edF/\hbar$ is the Bloch frequency. This latter is the second characteristic frequency of the system, which is overlooked by the effective mass approximation. The hopping matrix elements $J_x$ and $J_y$ in the Hamiltonian (\ref{6}) can be related via eq.~(\ref{3}) to the parameters of the original system as $J_{x,y}=\hbar^2/d^2 M^*_{x,y}$.  In what follows, to simplify equations, we shall set $e$, $c$, $\hbar$, $M$ and $d$ to unity. The magnetic period is then given by $\lambda = 1/\alpha$, the Bloch frequency becomes $\omega_B=F$, the cyclotron frequency
\begin{equation}
\label{6a}
\omega_c=2\pi\alpha(J_xJ_y)^{1/2}  \;,
\end{equation}
and the drift velocity
\begin{equation}
\label{6b}
v^*=F/2\pi\alpha  \;.
\end{equation}

The Hamiltonian (\ref{6}) with four parameters $J_x$, $J_y$, $\alpha$, and $\omega_B\equiv F$ defines the model being analyzed in the following sections.  More precisely, we are interested in wave-packet dynamics of the system (\ref{6}), which is governed by the Schr\"odinger equation for the amplitudes $\psi_{l,m}(t)$:
\begin{equation}
\label{7}
i\dot{\psi}_{l,m}=-\frac{J_x}{2}\left(\psi_{l+1,m}+\psi_{l-1,m}\right)
-\frac{J_y}{2}\left(e^{i(2\pi\alpha l-Ft)} \psi_{l,m+1}+e^{-i(2\pi\alpha l-Ft)} \psi_{l,m-1}\right) \;.
\end{equation}
Remark that the discrete index $l$ refers to the $x$ direction, while $m$ labels the $y$ direction, parallel to the electric field.

\subsection{Semiclassical analysis of a one-dimensional reduced Hamiltonian}
\label{sec2c}

We begin our analysis of the system dynamics with considering a class of initial conditions with uniform probability density along the $y$ direction. This leads to a one-dimensional reduction of the quantum problem. In fact, in this case we may use the substitution
\begin{equation}
\label{7z}
\psi_{l,m}(t)= \frac{e^{i\kappa m}}{\sqrt{L_y}}  b_l(t) \;,\quad \kappa=\frac{2\pi}{L_y}k \;,
\end{equation}
which reduces the Schr\"odinger equation (\ref{7}) to the following 1D equation for the amplitudes $b_l(t)$:
\begin{equation}
\label{7a}
i\dot{b}_l=-\frac{J_x}{2}(b_{l+1}+b_{l-1}) -J_y\cos(2\pi\alpha l+\kappa-Ft)b_l  \;.
\end{equation}
The substitution (\ref{7z}) assumes periodic boundary conditions, of period $L_y$, that eventually tends to infinity. Equation (\ref{7a}) can be rewritten in the more familiar form of the Schr\"odinger equation for the driven Harper Hamiltonian
\begin{equation}
\label{8}
\widehat{H}_{1D}=-J_x\cos\hat{p}-J_y\cos(2 \pi \alpha x+\kappa-Ft) \;.
\end{equation}
In this paper we shall mostly limit ourselves to the case $\alpha\ll 1$, which can be considered as a semiclassical limit for Harper-like Hamiltonians \cite{harpr}. Thus, we can appeal to the quantum-classical correspondence to facilitate the physical understanding of the problem. Letting the operators $\hat{x}'=2 \pi \alpha x$ and $\hat{p}'=-i2 \pi \alpha \; {\rm d}/{\rm d}x'$ be considered as classical canonical variables yields the the classical driven Harper Hamiltonian
\begin{equation}
\label{8b2}
 H_{cl}=-J_x\cos(p')-J_y\cos(x' - Ft) \;.
\end{equation}
\begin{figure}
\center
\includegraphics[width=11.5cm, clip]{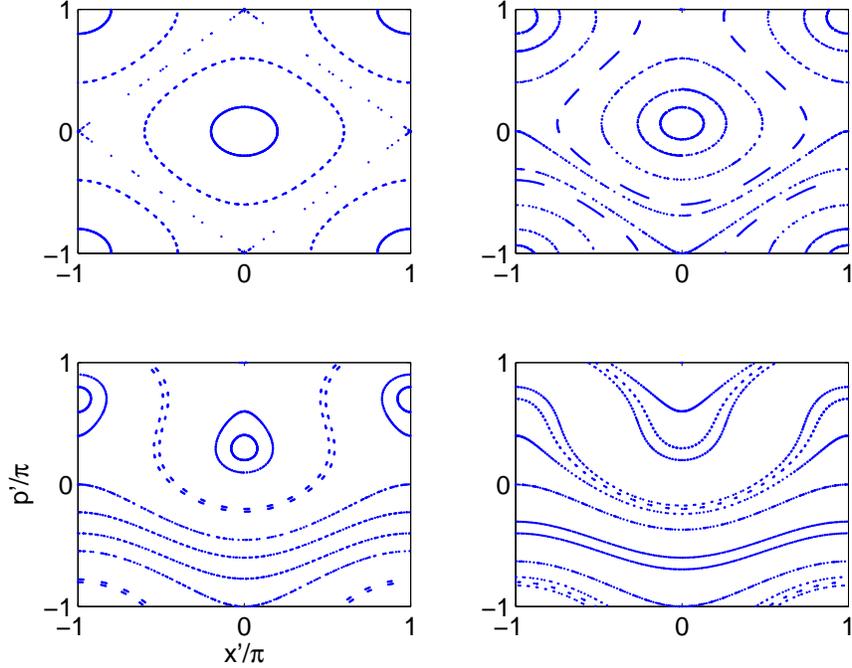}
\caption{(color online). Stroboscopic map of the driven Harper for $J_x=J_y=1$ and $F=0$, a small $F\ll F_{cr}$ and moderate  $F$ below and above $F_{cr}$.}
\label{fig1}
\end{figure}

\begin{figure}[t!]
\center
\includegraphics[width=11.5cm, clip]{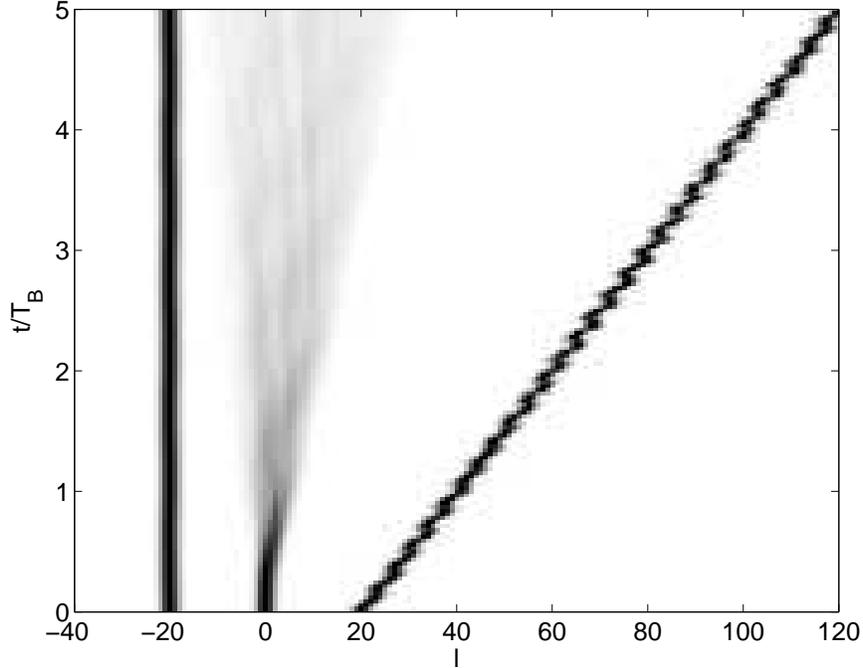}
\caption{Space--time plot of the dynamics of a wave packet initially localized about the origin for $J_x=J_y=1$, $\kappa=0$, $\alpha=1/20$ and $F=0,0.5,0.1$ (left, center, right). To increase visibility, the packets for $F=0$ and $F=0.1$ have been shifted by $\mp 20$ lattice sites. Wave-functions amplitudes $|b_l(t)|^2$ are reported in grey scale, black maximum. The time axis is measured in units of the Bloch period $T_B = {2 \pi}/{\omega_B}$ and we have arbitrarily set $T_B = 2 \pi$ for null electric field. For $F=0.1$ the particle is captured into the nonlinear resonance and transported across the lattice, traveling one magnetic period $\lambda$ in one Bloch period $T_B$.}
\label{fig2}
\end{figure}

Figure \ref{fig1} shows the Poincare' surfaces of section for the classical motions. Phase trajectories near the stationary point $(p',x')=(0,0)$ in the panel (a), which refers to the case $F=0$, encircle this point with the cyclotron frequency  (\ref{6a}) and are thus associated with the Landau states in the effective mass approximation. For weak field intensity $F$, these trajectories are captured into the nonlinear resonance, which is seen as the stability island in Fig.~\ref{fig1}(b-c), and are transported with the drift velocity $v^*$.  The stability island, however, shrinks with increasing $F$ and completely disappears when $|F|>F_{cr}$, where
\begin{equation}
\label{9}
F_{cr}=2\pi\alpha J_x \;.
\end{equation}
Thus the drift is possible only if $|F|<F_{cr}$. Translated into the quantum language this means that the Landau state can be transported only if the condition (\ref{9}) is satisfied.
To support this statement, in Fig.~\ref{fig2} we show the dynamics of a localized wave packet associated with the ground Landau state in the effective mass approximation, given by the initial condition $b_l(t=0)\sim\exp(-\pi\alpha l^2)$. A nice, directed transport is observed only in a weak field regime, while in the strong field regime the wave packet spreads almost symmetrically both directions. Remark that these results refer to the one-dimensional model (\ref{7a}). We will describe  wave-packet spreading in more detail in Sec.~\ref{sec4a}.

We also note that eq.~(\ref{9}) removes a seemingly illogical consequence of eq.~(\ref{6b}), that predicts an infinite drift velocity for null magnetic field. As a matter of facts, when the magnetic field intensity $B$ tends to zero, the transporting island in Figure \ref{fig1} disappears. Thus, eq.~(\ref{6b}) for the drift current implicitly assumes that $F < 2 \pi \alpha J_x$.

To conclude this section we note that directed transport in the weak field regime (i.e., the drift) takes place not only for states associated with trajectories near $(p',x')=(0,0)$, but also for states associated with trajectories near $(p',x')=(\pi,\pi)$ in Fig.~\ref{fig1}(a). These are high-energy counterparts of the low-energy (Landau) states of the Harper Hamiltonian, related by the transformation $b_l^{(high)}=(-1)^l b_l^{(low)}$.

\section{Quantum analysis of the 2D system}
\label{sec3}
We now move to the analysis of the two-dimensional tight--binding Hamiltonian. By using the Kramers--Hennenberger transformation, $\psi_{l,m}(t) \rightarrow \psi_{l,m}(t)\exp(-iFmt)$, we can reduce the Hamiltonian (\ref{6}) to the form
\begin{equation}
\label{c2}
(\widehat{H}_{tb} \psi)_{l,m} = -\frac{J_x}{2}\left(\psi_{l+1,m}+\psi_{l-1,m}\right)
-\frac{J_y}{2}\left(e^{i2\pi\alpha l} \psi_{l,m+1}+e^{-i2\pi\alpha l} \psi_{l,m-1}\right)
+Fm \psi_{l,m} \;.
\end{equation}
This can also be seen as the tight-binding approximation to the Hamiltonian
\begin{displaymath}
\widehat{H}=\frac{1}{2m}\left(\hat{\bf{p}}-\frac{e}{c}{\bf A}\right)^2+V(x,y) +eFy \;,
\end{displaymath}
where the vector potential  ${\bf A}=B(0,x,0)$ is used instead of that in eq.~(\ref{1a}). They describe the same physical situation.

\subsection{Landau-Stark states}
\label{sec3a}
One gets important insight into the dynamics by analyzing the energy spectrum of the Hamiltonian (\ref{c2}), $(\widehat{H}_{tb} \psi)_{l,m} = E \psi_{l,m}$. By assuming periodicity in the $x$ direction, of an arbitrary (large) period $L_x$, and eventually letting $L_x$ tend to infinity, we can write the eigenfunctions in the form
\begin{equation}
\label{c3}
\psi_{l,m}= \frac{e^{i\kappa l}}{\sqrt{L_x}} c_m e^{-i2\pi\alpha lm} ,
\end{equation}
where $\kappa$ is free to vary in the interval $[0,2 \pi]$. [Please notice the difference between eqs. (\ref{c3}) and (\ref{7z}).] This reduces the spectral problem to the one-dimensional problem:
\begin{equation}
\label{c4}
-\frac{J_y}{2}(c_{m+1}+c_{m-1}) + [Fm - J_x\cos(2\pi\alpha m-\kappa)]c_m =E c_m \;.
\end{equation}
Labeling the eigen-solutions of (\ref{c4}) by the discrete index $\nu$, so that $E = E_{\nu}(\kappa)$ and $c_m = c_m ^{\nu}(\kappa)$ and inserting these values in eq.~(\ref{c3}), we construct the eigenfunctions of (\ref{c2}) in the form of the {\em Landau--Stark states}:
$|\Psi_{\nu,\kappa}\rangle=\sum_{l,m} \psi_{l,m}^{(\nu,\kappa)} |l,m\rangle$.

\begin{figure}
\center
\includegraphics[width=11.5cm, clip]{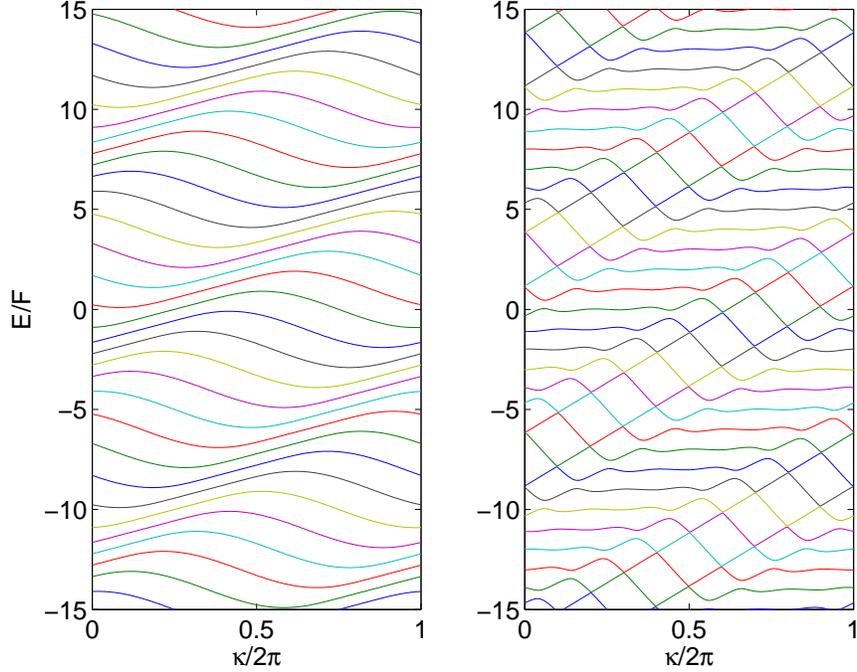}
\caption{(color online). A fragment of the energy spectrum of (\ref{c2}) for $J_x=J_y=1$, $\alpha=1/10$, and $F=1$ (left) and $F=0.3$ (right).}
\label{fig3}
\end{figure}

For $J_x=0$ the spectrum of (\ref{c4}) consists of a ladder of Bloch bands of zero width, separated by the Stark energy $F$. For $J_x\ne0$ these bands acquire a finite width $\sim 2J_x$ and for any fixed $\kappa$ the spectrum is a modulated Wannier-Stark ladder. A fragment of the spectrum covering three modulation periods is shown in Fig.~\ref{fig3} for two values of the field intensity $F$. Note that for small $F$ a pattern of straight lines emerges in the picture. The eigenstates belonging to this lines (which always appear in pairs) are associated with two transporting islands in the classical approach. We shall discuss them in more details in the next subsection.

It follows from eq.~(\ref{c3}) that Landau-Stark states are extended states in the $x$ direction. Moreover, they are current states, {\em i.e.} the mean value of the current operator  $\hat{v}$,
\begin{equation}
\label{c5}
\hat{v}=\frac{J_x}{2i}\left(\sum_{l,m} |l+1,m\rangle\langle l,m| -h.c.\right) \;,
\end{equation}
is non-zero for almost all Landau-Stark states $|\Psi_{\nu,\kappa}\rangle$.  This value, $v_\nu(\kappa)$, can be calculated via eq.~(\ref{c3}):
\begin{equation}
\label{c6}
v_\nu(\kappa)=\langle \Psi_{\nu,\kappa}|\hat{v}| \Psi_{\nu,\kappa}\rangle
=\sum_m |c_m^{\nu}(\kappa)|^2 \sin(2\pi\alpha m -\kappa)  \;.
\end{equation}

\begin{figure}
\center
\includegraphics[width=11.5cm, clip]{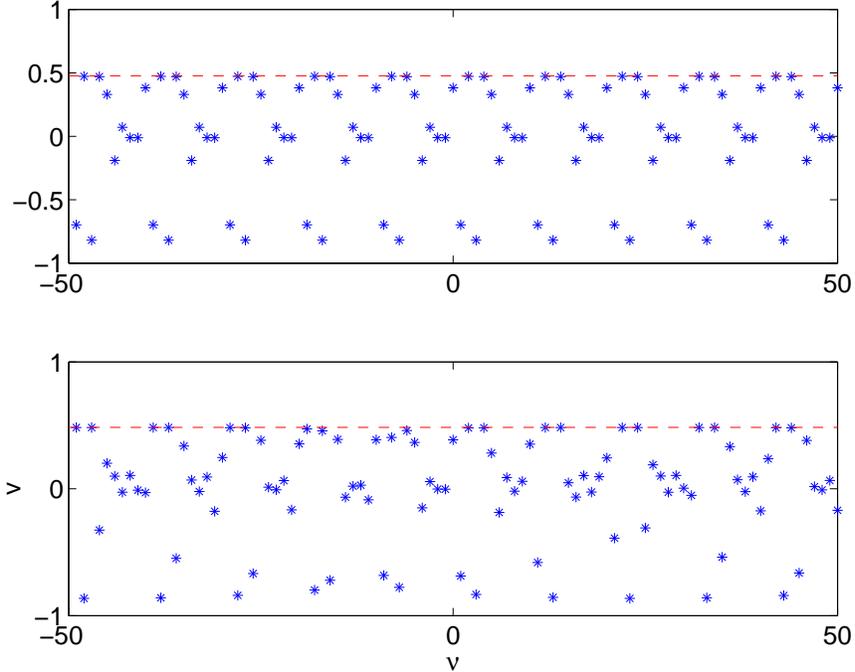}
\caption{(color online). The mean current (\ref{c5}) carried by Landau-Stark states  $|\Psi_{\nu,\kappa}\rangle$ with $\kappa=0.1$. Parameters are  $J_x=J_y=1$, $F=0.3$, $\alpha=1/10$ (upper panel) and $\alpha=1/10.1417$ (lower panel). The dashed line corresponds to the drift velocity $v^*$.}
\label{fig4}
\end{figure}

The upper panel in Fig.~\ref{fig4} shows the mean current for a set of one-hundred states $|\Psi_{\nu,\kappa}\rangle$ with $\kappa=0.1$ for the same parameters of Fig.~\ref{fig3}(b). It is seen that the Landau-Stark states carry currents of different magnitude and sign. However, completeness of the basis of eigenstates implies the sum rule
\begin{equation}
\label{c7}
\sum_\nu v_\nu(\kappa)=0 \;.
\end{equation}
%

A remark about commensurably of the magnetic and lattice periods is in turn. In Fig.~\ref{fig3} the spectrum $E_\nu(\kappa)$ is periodic along the energy axis because $\alpha$ is rational. This periodicity is replaced by quasi-periodicity when $\alpha$ is irrational. As a consequence, the same happens also for $v_\nu(\kappa)$, as can be seen clearly in Fig.~\ref{fig4}(b). However, Landau-Stark states are still extended functions in the $x$ direction and the summation rule (\ref{c7}) also holds.  Thus, irrationality of the value $\alpha$ does not change properties of the system in a crucial way. This differs from the case of vanishing electric field, where the eigenstates are localized functions for irrational $\alpha$ if $J_x>J_y$ \cite{Aubr80}.

\subsection{Transporting states}
\label{sec3b}
In this subsection we describe a procedure for constructing localized wave packets, which propagate across the lattice with the drift velocity velocity $v^*$ in eq.~(\ref{6b}) without changing their shapes.

\begin{figure}
\center
\includegraphics[width=11.5cm, clip]{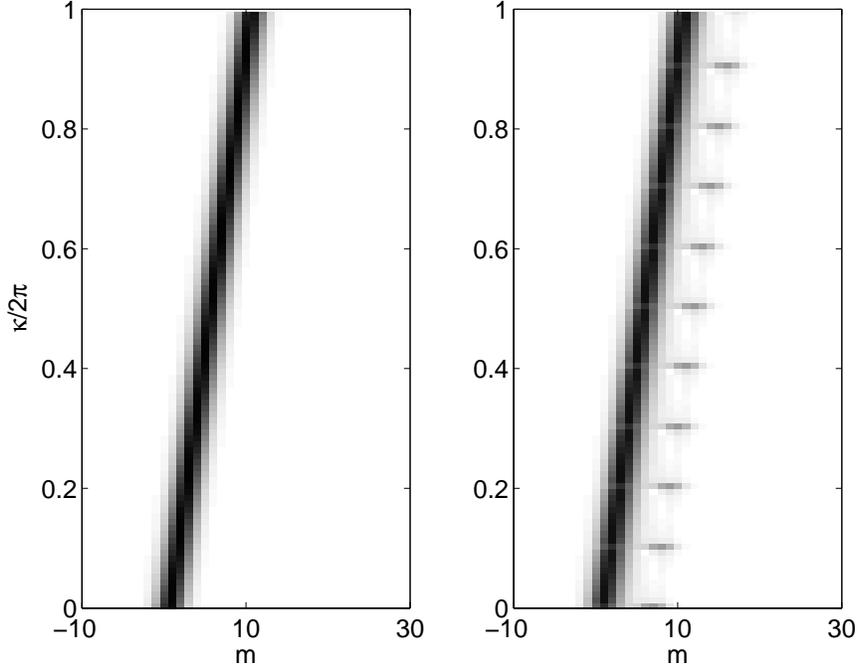}
\caption{Absolute values of the expansion coefficients $c_m$, calculated on the basis of the exact equation (\ref{c4}) and the approximate equation (\ref{c8}). Parameters are  $J_x=J_y=1$, $F=0.3$, and $\alpha=1/10$.}
\label{fig5}
\end{figure}

First of all we note that for $|F|<F_{cr}$ the potential energy term in eq.~(\ref{c4}), $V(m)=Fm-J_x\cos(2\pi\alpha m-\kappa)$, has local minima and maxima. These local extrema can support a number of well localized states. The exact number of these states depends on the system parameters and coincides with number of straight lines in the energy spectrum. In first order approximation they can be found by expanding $V(m)$ in the vicinity of the extremum.  Without loss of generality, let us consider the minimum at $m=0$. This expansion reduces (\ref{c4}) to the following equation,
\begin{equation}
\label{c8}
-\frac{J_y}{2}(c_{m+1}+c_{m-1})
+ \left[Fm+\frac{J_x}{2}(2\pi\alpha m-\kappa)^2 \right]c_m = E c_m
\end{equation}
(the constant term is omitted), which  is a Mathieu-type equation whose properties are well-known.  As an example, Fig.~\ref{fig5} compares the localized state, which is calculated using the exact equation (\ref{c4}) with that following from the approximate equation (\ref{c8}) for the parameters of Fig.~\ref{fig3}(b). The two states well coincide with the exception of ten points in the $\kappa$ axis, in coincidence of which [see Fig.~\ref{fig3}(b)] the spectrum shows avoided crossings.

\begin{figure}
\center
\includegraphics[width=11.5cm, clip]{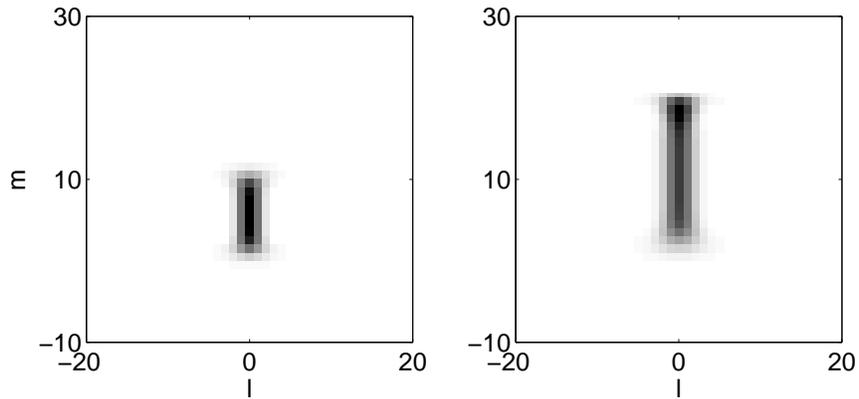}
\caption{Grey scale representation of transporting states. Parameters are  $J_x=J_y=1$, $F=0.1$, and $\alpha=1/10$ (left) and $\alpha=1/20$ (right).}
\label{fig6}
\end{figure}

Ignoring avoided crossings [i.e., using eq.~(\ref{c8}) instead of eq.~(\ref{c4})] and tracking the energy $E_\nu(\kappa)$ along the straight lines in Fig.~\ref{fig3}(b) we can construct a localized state by integrating extended Landau-Stark states over the quasimomentum,
\begin{equation}
\label{c9}
\psi_{l,m}= e^{-i2\pi\alpha lm} \int_0^{2\pi} e^{i\kappa l} c_m(\kappa) {\rm d}\kappa \;.
\end{equation}
It follows from eq.~(\ref{c8}) that the time evolution of this state is defined by the trivial shift,
\begin{equation}
\label{packet}
\psi_{l,m}(t)=\psi_{l-v^*t,m}(0) \;,
\end{equation}
where $v^*$ is the drift velocity.  Needless to say, transporting states can be constructed for any local minimum or maximum of $V(m)$, {\em i.e.} for any straight line in the spectrum. Examples of transporting states for $F=0.1$ and two different values of $\alpha$ are given Fig.~\ref{fig6}. It is seen that the states (\ref{c9}) are well localized in both spatial directions, the characteristic localization length along the $y$ axis being defined by the magnetic period $\lambda = 1/\alpha$.

\begin{figure}
\center
\includegraphics[width=11.5cm, clip]{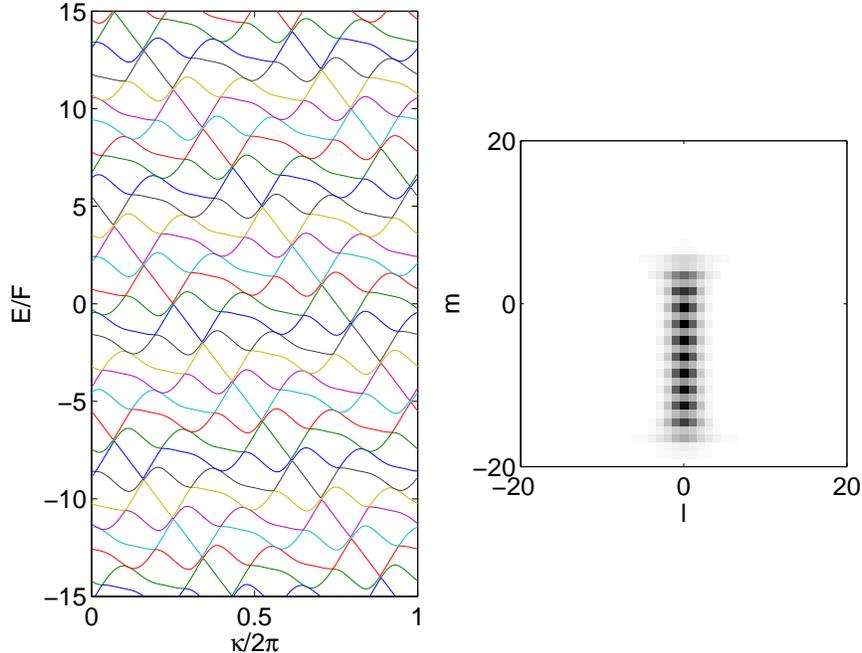}
\caption{(color online). A fragment of the energy spectrum of the system (left) and  the 'exotic' transporting state (right), which transports particles in the counter-intuitive direction. The system parameters are $J_x=J_y=1$, $F=0.1$ and $\alpha=1/(2.2)$.}
\label{fig7}
\end{figure}

We mentioned above that transporting states are associated with transporting islands in the semiclassical approach, which is justified if $|\alpha|\ll 1$. An advantage of the Landau-Stark states formalism is that one can make predictions also for $|\alpha|\sim 1/2$, i.e., in the deep quantum regime. (Without a loss of generality one may consider $-1/2<\alpha\le 1/2$.)   In particular, for $\alpha\approx1/2$ we have found exotic transporting states, which have no classical analogue. An example is given in the left panel of Fig.~\ref{fig7}, which shows the band spectrum $E_\nu(\kappa)$ for $\alpha=1/(2.2)$. Using the procedure described above one can construct localized transporting states for any straight line in the spectrum, see right panel in Fig.~\ref{fig7}. Note that, while in Fig.~\ref{fig3}(b) the slope of the straight lines is positive, in Fig.~\ref{fig7}(a) it is negative: therefore the depicted localized state transports quantum particles in the opposite direction to what (classically) expected by the electric-magnetic field geometry.

In closing this section we need to stress again that we ignored avoided crossings when constructing the transporting states. The presence of these avoided crossings leads to back-scattering effects: the traveling packet in eq.~(\ref{packet}) emits tiny packets, which propagate in the opposite direction.  Thus, strictly speaking, the perfect transport of a localized wave packet in a lattice is impossible.

\section{Wave--Packet spreading}
\label{sec4}
The band spectrum of the Landau-Stark states and the summation rule (\ref{c7}) imply that the time evolution of a generic localized wave packet is a ballistic spreading along both the positive and the negative $x$ direction, with the obvious exception of specific initial conditions such as the transporting states. In this section we study wave--packet spreading in some detail, extending the analysis also to the case of disordered lattices. To construct a generic initial wave packet, we shall superimpose random phases to a wide Gaussian envelope of width $\sigma_0\gg 1/\alpha$. Note that by averaging the results over such random phases we can mimic the dynamics of an incoherent wave packet.

\subsection{One-dimensional system}
\label{sec4a}

To have a reference model for the phenomena under investigation, we first analyze the dynamics in the one-dimensional system (\ref{8}).
The simplest situation is obtained for large $F\gg F_{cr}$, where the classical phase--space trajectories are nearly straight lines parallel to the $x$ axis. In this case we can safely neglect the potential energy term and hence assume that the spreading motion coincides with that in the absence of driving. For an incoherent wave packet this implies that the square root of the second spatial moment, $\sigma(t)$, is
\begin{equation}
\label{d1}
\sigma(t)=\sqrt{\sigma_0^2+J_x t^2/2} \;.
\end{equation}
Equation (\ref{d1}) gives an upper boundary to the rate of ballistic spreading of a localized wave packet.
Decreasing $F$ the rate of spreading also diminishes. At the same time, the spreading becomes asymmetric, a fact especially evident for $F<F_{cr}$. In this regime a large part of the packet moves in the positive direction with the drift velocity $v^*$, while a smaller part moves in the negative direction with velocity $|v|>v^*$. Note that this asymmetric spreading does not imply a preferable transport in either direction, since the first moment of the position of the packet is null. Fig.~\ref{fig9}  below provides a numerical confirmation of this fact. It is obtained by averaging over the random phases of the initial wave packet.

\begin{figure}
\center
\includegraphics[width=11.5cm, clip,angle=0]{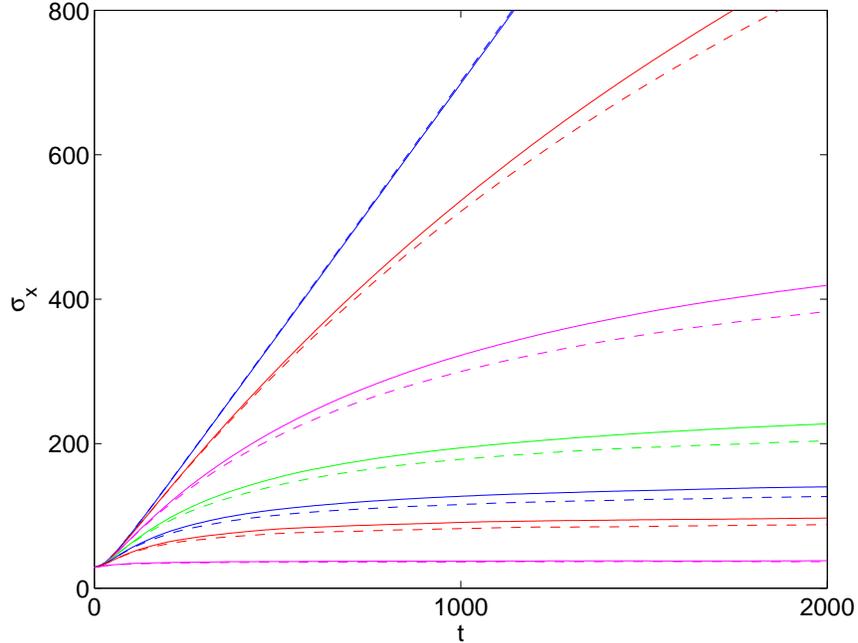}
\caption{(color online). Wave-packet width as a function of time for $J_x=J_y=1$, $\alpha=1/10$, $F=3$ and increasing strength of the disorder $\epsilon=0,0.1,0.2,03,0.4,0.5,1$, (color labelled as blue, red, magenta, green, blue, red, magenta, respectively) from top to bottom. The continuous lines correspond to the 2D model, while the dashed lines to the one--dimensional model.}
\label{fig8a}
\end{figure}

Next we study the effect of on-site disorder, which we mimic by adding to the 1D Hamiltonian in eq.~(\ref{8}) random on-site energies $\epsilon_n$, $|\epsilon_n| \le \epsilon/2$. The dashed lines in Fig.~\ref{fig8a} show $\sigma(t)$ versus $t$ for increasing values of the disorder, at fixed $F=3$. Here, and in the following, we also perform an average over disorder realizations. As expected, the ballistic spreading is suppressed by the disorder. Moreover, for large times the spreading is completely stopped by the Anderson localization. In this case the spatial distribution of the average packet, $P(l,t)=\langle |\psi_l(t)|^2 \rangle$ approaches the exponential distribution,
\begin{equation}
\label{d2}
P(l,t) \sim \exp(-|l|/{\cal L}) \;,
\end{equation}
which is a hallmark of Anderson localization. Let us also note that the value $F=3$ chosen in Fig.~\ref{fig8a} is large enough to approach the free expansion regime, eq.~(\ref{d1}), for  $\epsilon=0$. Thus we expect that for $\epsilon\ne0$ the functional dependence of the localization length ${\cal L}={\cal L}(\epsilon)$ in eq.~(\ref{d2}) might coincide with that of the standard Anderson model.

\begin{figure}
\center
\includegraphics[width=11.5cm, clip,angle=0]{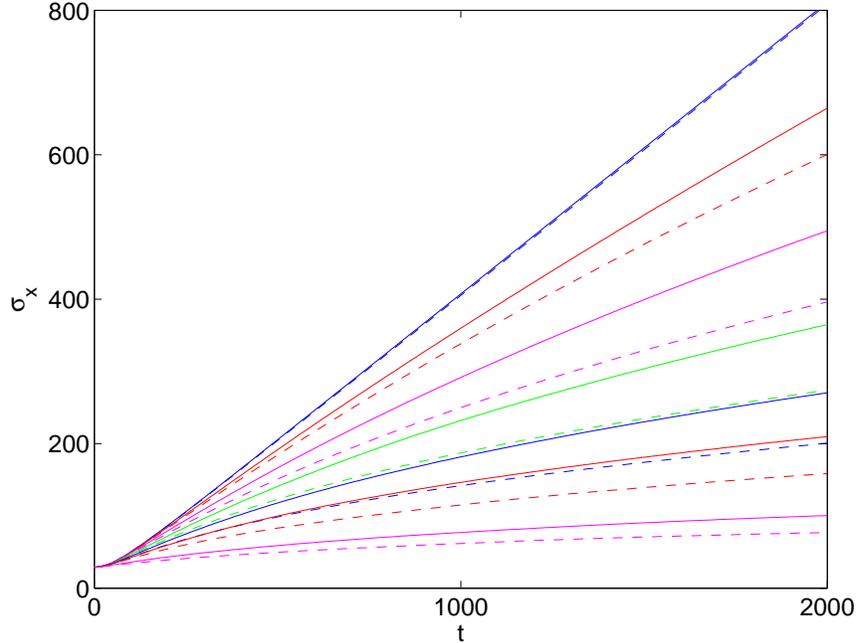}
\caption{(color online). The same as in the previous figure, now for $F=0.3$.}
\label{fig8b}
\end{figure}

The case $F<F_{cr}$, shown in Fig.~\ref{fig8b} (dashed lines), appears to be more complicated. Here the regime of ballistic spreading changes to the regime of Anderson localization through an intermediate regime of diffusive spreading, where the spatial distribution function of the wave packet grows approximately like the diffusion law,
$
P(l,t) \sim \exp(-l^2/Dt). \;
$
At  $F=0.3$ this regime is clearly observed in the numerical simulations for disorder strengths  around $\epsilon \sim 0.5$. To see that this is actually an intermediate regime, ({\em i.e.} to see a sign of saturation) we will perform a more sophisticated analysis, in the next section.

\subsection{Two-dimensional system}
\label{sec4b}

As in the 1D case we consider a Gaussian incoherent packet as initial condition for the motion, of width $\sigma_0^x$ and $\sigma_0^y$ in the two directions; the former is chosen to coincide with the width of the initial condition in the one--dimensional model. For vanishing disorder the packet spreads ballistically in the $x$ direction but remains localized in the $y$ direction. The Landau-Stark localization length ${\cal L}_y$ in the $y$ direction is in fact:
\begin{equation}
\label{d4}
{\cal L}_y \approx \left\{
\begin{array}{ccc}
\max (1,2J_y/F) &,& F>F_{cr}\\
1/\alpha &,& F<F_{cr}
\end{array} \right.
\end{equation}
The estimate (\ref{d4}) is obtained on the basis of the 1D Wannier-Stark problem (\ref{c4}) and assumes $\alpha\ll 1$ and $J_x=J_y$ (note however \cite{remark2}). It follows from eq.~(\ref{d4}) that the localization length increases from unity to one magnetic period when $F$ is decreased. We also mention that in the under-critical regime, $F<F_{cr}$, the estimate  ${\cal L}_y\approx1/\alpha$ gives the maximal localization length, while the minimal localization length scales as  ${\cal L}_y\sim1/\sqrt{\alpha}$.  Thus in this regime different Landau-States have different localization lengths ${\cal L}_y$, in the interval $1/\sqrt{\alpha}<{\cal L}_y<1/\alpha$.

\begin{figure}
\center
\includegraphics[width=11.5cm, clip]{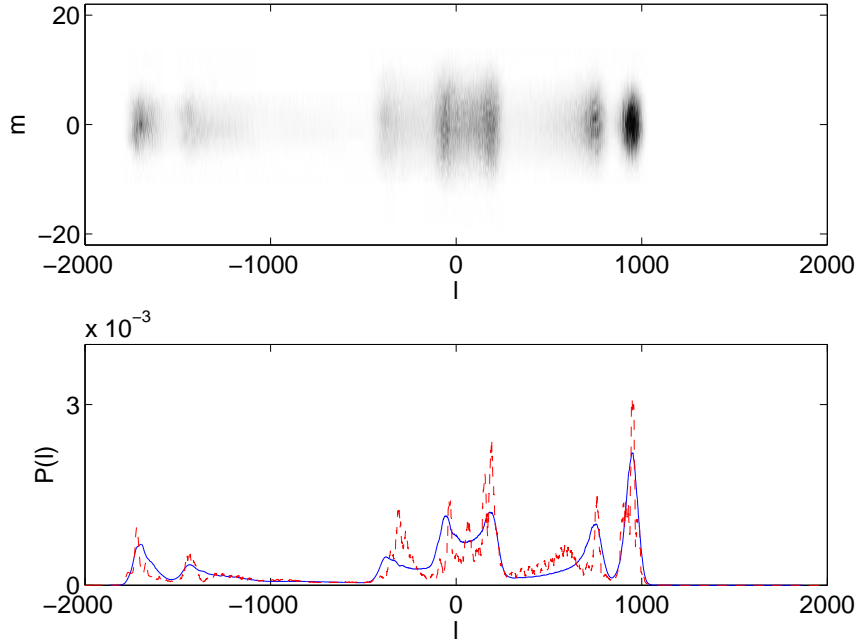}
\caption{(color online). Upper panel: Population of the lattice sites at the end of numerical simulation. Lower panel compares integrated population of the lattice sites along the $x$ direction (solid line) with prediction of the 1D model (\ref{7}), dashed line.}
\label{fig9}
\end{figure}

We can now show the results of the analysis of the full, two-dimensional system (\ref{c2}). The upper panel in Fig.~\ref{fig9} depicts the averaged 2D wave-packet $\langle |\psi_{l,m}(t)|^2 \rangle$ at $t=2000$ for the parameters of Fig.~\ref{fig3}(b) (note the different scale of the $y$ and $x$ axis). It is seen that the packets splits in several sub-packets, of which the rightmost moves with the drift velocity. This sub-packet is associated with the straight lines in the energy spectrum, i.e., with the transporting states. At the opposite end, the leftmost sub-packet is associated with the part of the spectrum where the energy bands have minimal negative derivative. Intermediate packets are also observed, so that one can easily predict speed and position of wave-packets by analyzing the energy spectrum.

The lower panel in Fig.~\ref{fig9} compares the projected wave-packet in the $x$ direction, $\sum_m |\psi_{l,m}|^2$, with the corresponding distribution predicted by the 1D model (\ref{7a}). A reasonable correspondence is noticed, which strengthen the physical significance of the one--dimensional reduction. This correspondence is even more precise if one compares integrated characteristics like the wave-packet widths, which we define as
\begin{displaymath}
\sigma(t)= (\sum_{l} l^2 |c_{l}(t)|^2 )^{\frac{1}{2}}  \quad \mbox{and} \quad
\sigma(t)=\ (\sum_{l,m} l^2 |\psi_{l,m}|^2 )^{\frac{1}{2}}
\end{displaymath}
in the 1D and 2D cases, respectively. This comparison is done in Fig.~\ref{fig8a} and Fig.~\ref{fig8b} for $F=3$ and $F=0.3$, respectively. It is seen that the values for $\epsilon=0$ practically coincide. To the contrary, for non-zero amplitude of the disorder, the data of the one--dimensional model are consistently smaller than the other, supporting the almost obvious observation that in the 2D model Anderson localization is ``less effective''. To turn this observation into rigorous theory is the goal of further investigation.

\begin{figure}
\center
\includegraphics[width=11.5cm, clip,angle=-90]{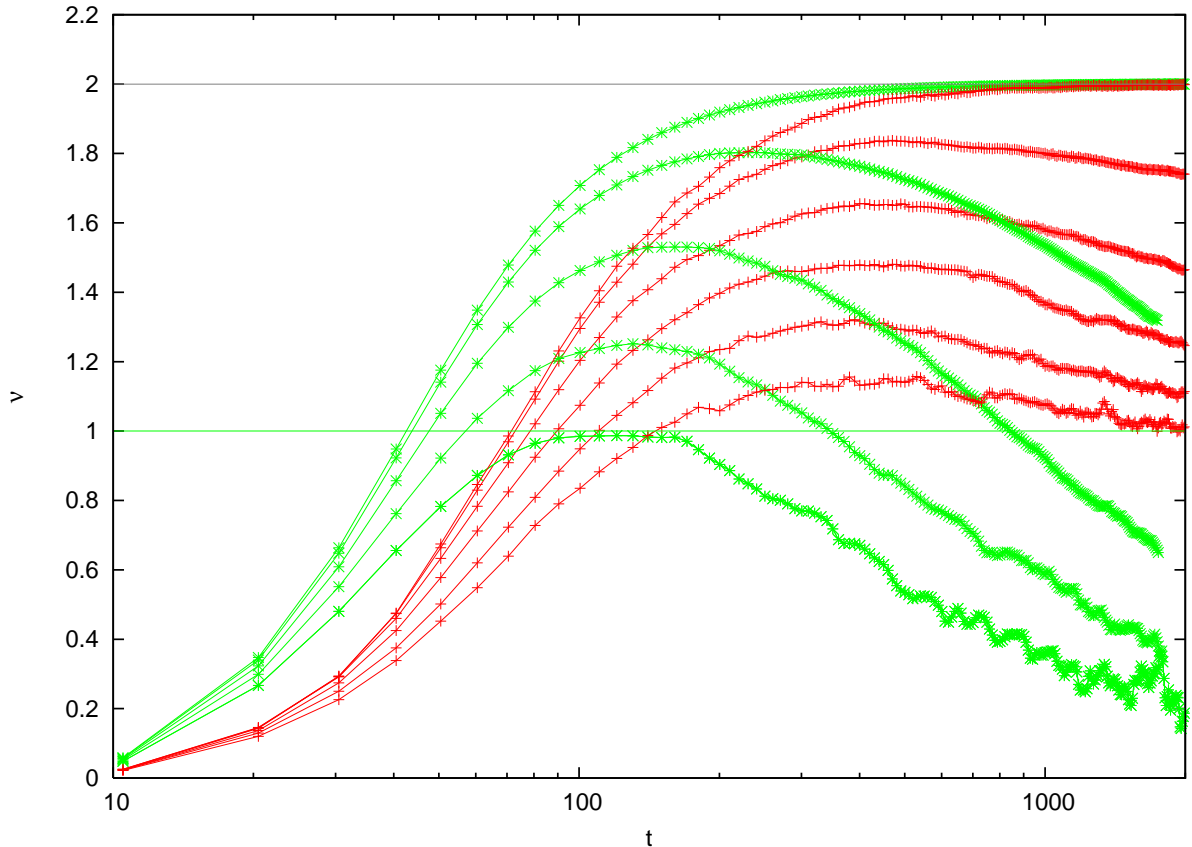}
\caption{(color online). Local exponents $\nu(t)$ in the growth of the second moment of wave--packets as a function of time, computed on the data of Fig. ~\ref{fig8a} and Fig.~\ref{fig8b}. Labels: $F=0.3$, green lines, asterisks; $F=3.$ red line, crosses. In both cases $\epsilon=0,.1,.2,.3,.4$ (top to bottom curves). The horizontal lines at $\nu=1$ and $\nu=2$ mark the linear and quadratic increase of $\sigma^2$.}
\label{figslop}
\end{figure}

Here, we only perform a simple analysis. From the data of Fig. ~\ref{fig8a} and Fig.~\ref{fig8b} one might think that anomalous diffusion is taking place in the dynamics of the wave--packet, as in well known examples of quantum intermittency (see \cite{gio} and references therein). Indeed, this is not the case. In fact, if we compute the local exponent $\nu(t)$ as the slope of the curve $\log(\sigma^2(t))$ versus $\log t$:
\begin{displaymath}
\nu(t)=  \frac{d \log( \sigma_x^2(t) )}{d \log t} \; ,
\end{displaymath}
we obtain Fig. \ref{figslop}, that sums up the data for $F=0.3$ and $F=3.0$. While slopes for $\epsilon=0$ tend clearly to the value $\nu = 2$, indicating linear motion, the other curves, after reaching a maximum, show a clear tendency to decrease. (The initial increase can be easily explained as an effect of the wide amplitude of the initial wave-packet.) The successive decrease of the local exponent $\nu(t)$ is the signature of Anderson localization. In turn, this can be justified by noting that Stark localization in the $y$ direction, eq.~(\ref{d4}), renders the system quasi one-dimensional, effectively bounding the motion to a strip, the larger the value of the electric field amplitude $F$, the narrower this strip.

\section{Conclusions}
We have studied the wave-packet dynamics of a quantum particle in a square lattice in the presence of (real or artificial) electric and magnetic fields, in the tight-binding approximation. In this approximation the Hamiltonian of this system contains five physical parameters: the hopping matrix elements $J_{x,y}$, the components of the electric field vector  $F_{x,y}$, and the magnetic field flux quanta through a unit cell $\alpha$,  (with $|\alpha|\le 1/2$). In this paper we have considered mainly the case  $J_y=J_x\equiv J$, but we do not expect significant differences to appear in the more general situation. More importantly, we have assumed that the electric field is parallel to one of the lattice axes which, in fact, is a rather specific situation. In spite of this limitation the system is found to display very rich dynamics, varying from ballistic spreading to directed transport. A laboratory application of this model can consist of cold atoms in a 2D optical lattice,  subject to artificial electric and magnetic fields. In this experiment, typical initial conditions correspond to a coherent (a BEC of atoms) or incoherent (a thermal cloud of atoms) 2D wave packet, localized over several lattice sites. Our findings apply directly to this experimental situation.

We have discovered two main regimes for wave-packet dynamics under these geometric conditions. These regimes are determined by the relation between two dimensionless parameters -- the parameter $\alpha$, which is proportional to magnetic field magnitude and the parameter $\beta$, which is  proportional to the electric field magnitude and is defined as the ratio of the Stark energy to the hopping matrix element.

For vanishing magnetic field, $\alpha=0$, the electric field induces Bloch oscillations of the wave packet in the field direction, while in the orthogonal direction the wave packet spreads ballistically with a rate defined by $J$. Under the condition of Stark localization, $\beta \gg 1$,  this regime is also valid if $\alpha\ne0$. Note that the condition $\beta\gg 1$ implies that the amplitude of Bloch oscillations is smaller than one lattice site.

In the weak field regime, $\beta \ll 1$, the dynamics are richer and depend on the relative size between the parameters $\alpha$ and $\beta$. Namely, if $\beta>2\pi\alpha$  the wave-packet still spreads ballistically, although the spreading becomes asymmetric.  A qualitative change happens if $\beta<2\pi\alpha$. Note that in dimensional units this condition reads $\omega_B<\omega_c$, where $\omega_B$ and $\omega_c$ are the Bloch and cyclotron frequencies, respectively. Here, the wave packets splits into several sub--packets, of which the right--most moves with the drift velocity $v^*$. This sub-packet is supported by the transporting states, the explicit form of which is given in Sec.~\ref{sec3b}.

We have explained the observed dynamical regimes by analyzing the energy spectrum of the system, which can be easily calculated numerically. This spectrum has a band structure with specific band patterns, sensitive to the system parameters. Analyzing these patterns one can make predictions about the wave-packet dynamics even before performing the demanding simulation of the 2D Schr\"odinger equation.

In this work we have also derived and tested a 1D approximation to the original 2D  Schr\"odinger equation, which leads to the driven Harper Hamiltonian (\ref{8}). This latter is interesting in its own, because the driven Harper system is more easly realized in the lab than 2D lattices with artificial electric and magnetic fields \cite{Roat08}.  We have shown that the 1D model is capable to capture some important features of the 2D wave-packet dynamics.

Finally we have studied the effect of on-site disorder on the dynamics \cite{Bill08}.  It is found that disorder changes both the ballistic and transporting regimes into the regime of Anderson localization, which can be preceded by a diffusive regime. The observed diffusive spreading of the wave packet together with asymptotic Anderson localization indicates that on-site disorder converts the extended Landau-Stark states (which are the system eigenstates for vanishing disorder) into localized states with a non--trivial scaling law. This scaling law is a problem open to further investigation.

An additional open problem, which we reserve for future studies, is the case of an arbitrary direction of the electric field. It is known that Bloch oscillations of a quantum particle  on a 2D lattice changes drastically if the field vector is misaligned with respect  the lattice axis \cite{58}. Thus one may expect the quantum transport in the presence of a magnetic field to be also different.

\vspace*{5mm}
\noindent
{\it Acknowledgments}\\
A.K. acknowledges the generous support of the Cariplo Foundation of the Landau Network-Centro Volta, Italy, supporting his stay at Center for Nonlinear and Complex Systems of the University of Insubria at Como, where part of this work has been performed. G.M. acknowledges
the support of MIUR-PRIN project ``Nonlinearity and disorder in classical and quantum transport processes''.
The calculations in this paper have been produced on the CINECA parallel processor cluster SP6 thanks to the grant "Open quantum systems: quantum entropy and decoherence" for Progetti di Supercalcolo in Fisica della Materia, Project Key : giorgiomantica376572968172.


\begin{thebibliography}{10}
\bibitem{remark0}
The problem of Bloch oscillations in the presence of a magnetic field was addressed earlier with respect to semiconductor superlattices (see \cite{Cana96,Orli08}, for example). However, semiconductor superlattices are quasi one-dimensional lattices. Thus, in spite of similarities in terminology, we consider a completely different problem.
\bibitem{remark2}
For $J_y\ll J_x$ the estimate ${\cal L}_y=1/\alpha$ for the maximal localization length should be substituted by ${\cal L}_y=2(J_y/J_x)^{1/2}/\pi\alpha$ and, hence, ${\cal L}_y$ is smaller than one magnetic period. In the opposite limit $J_y\gg J_x$ the maximal localization length is larger than one magnetic period. However, it cannot exceed ${\cal L}_y=2J_y/F$, which is the localization length of the Wannier-Stark states.
\end{thebibliography}
\end{document}